\documentclass[10pt,conference]{IEEEtran}
\IEEEoverridecommandlockouts
\IEEEpubid{\makebox[\columnwidth]{\hfill \copyright~2018 IEEE}}

\usepackage[parfill]{parskip}

\usepackage{amsmath}
\usepackage{amssymb}
\usepackage{graphicx}
\usepackage{times}
\usepackage{caption}
\usepackage{xcolor}

\allowdisplaybreaks

\begin{document}
\pagenumbering{gobble}

\title{\textbf{\huge Multiplication with Fourier Optics\\[-.3ex] Simulating 16-bit Modular Multiplication}}

\author{\IEEEauthorblockN{\large Abigail N. Timmel}
\IEEEauthorblockA{Booz Allen Hamilton\\308 Sentinel Drive\\
Annapolis Junction, MD 20701\\
Email: {\tt timmel.abigail@bah.com}}
\and
\IEEEauthorblockN{\large John T. Daly}
\IEEEauthorblockA{Laboratory for Physical Sciences\\ 5520 Research Park Drive\\
Catonsville, MD 21128\\
Email: {\tt jtdaly3@lps.umd.edu}}}

\maketitle

\begin{abstract}
This paper will describe a simulator developed by the authors to explore the design of Fourier transform based multiplication using optics. Then it will demonstrate an application to the problem of constructing an all-optical modular multiplication circuit. That circuit implements a novel approximate version of the Montgomery multiplication algorithm that enables the calculation to be performed entirely in the analog domain. The results will be used to corroborate the feasibility of scaling the design up to 16-bits without the need for analog to digital conversions at intermediate steps.
\end{abstract}


\IEEEpeerreviewmaketitle

\section{Background}
Optical computing, performing calculations using photons instead of electrons, is an idea that has been around for decades~\cite{sawchuk84, psaltis86, feitelson88, touch17}. Like many competing non-CMOS technologies however, it was ultimately beat out by Moore's Law scaling of transistors. Optics, being fundamentally limited by the wavelength of light, cannot attain the density of transistors which, in 2017, achieved a 10nm process technology in commercial electronics~\cite{samsung}. Furthermore, as an analog technology{\color{black}, which is typically limited to 8-12 bits of resolution~\cite{hasler17},} optical computing does not compete with digital technologies in terms of computational accuracy. In spite of the hegemony of CMOS technology for computing, there continues to be active research in methods of leveraging optics for computation. With the recent slowing of Moore's Law scaling accompanied by advances in nanophotonics~\cite{koenderink15} and some initial commercial adoption of optical computing technologies~\cite{optalysys}, the research community has demonstrated renewed interest in exploring the role of optics in post-Moore's era computing~\cite{rebooting}.

A key strength in optical computing is the use of Fourier transforms to simplify convolution problems.  In particular, Fourier transforms can be used to turn multiplication from an $O(n^2)$ problem into an $O(n)$ one.  This is due to the distributive property: multiplying $(a + b) (c + d)$ gives the same result as adding $(ac + ad + bc + bd)$.  If we can separate out the result of the former, we get four multiplies by only performing one.  This suggests that any multiplication problem can be simplified by summing the digits of each $n$-digit factor and then multiplying.  Separating out the $2n$ multiplies in the result is still an issue, but we can resolve this by using a Fourier transform to sum the digits of our factors.  If we convert each factor into a function that takes on the value of each digit for some interval, the Fourier transform, effectively gives us the sum of the digits, each multiplied by a factor of $\mathrm{e}^{-i x y}$.
\begin{equation}
F(y) = \int_{-\infty}^{\infty} f(x) \mathrm{e}^{-i x y} \mathrm{d} x
\end{equation}

We can then do a point-wise multiply (multiply each point of one Fourier transformed function by the corresponding point on the other Fourier transformed function) to obtain the sum of products in each point of the result.  To separate out these products, we do the inverse Fourier transform:
\begin{equation}
f(x) = \int_{-\infty}^{\infty} F(y) \mathrm{e}^{i x y} \mathrm{d} x
\end{equation}
This gives us a function that represents the product of the numbers encoded in the original two functions.

The reason this technique is not widely used on digital computers is that Fourier transforms are expensive to compute~\cite{schonhage71, furer07}.  There is only an advantage for sufficiently large multiplication problems.  However, it can be shown that Fourier transforms can easily be achieved optically by using lenses and masks~\cite{chandran86, oel}. {\color{black} Optical correlators designed to perform multiplication, particularly vector-matrix multiplication, are not new~\cite{shaked07}. Neither are the optical simulations used to design such systems. However, the authors believe this to be the first time that such correlators have been demonstrated by simulation as a viable design for modular multiplication in optics. Therefore, the contributions of this work are the end-to-end simulation of a novel application of Fourier optics to the implementation of high precision arithmetic and a new approach to computing modular multiplication based on discrete convolution. }

\IEEEpubidadjcol

\section{Lens/Mask Configuration}
The equation that governs the electromagnetic field $U_1$ in a plane behind a mask is as follows~\cite{ocs130, goodman96, kim11}:
\begin{equation}
\label{field}
U_1(x_1,y_1) = \int_{-\infty}^{\infty} \int_{-\infty}^{\infty} U_0(x_0,y_0) \mathrm{e}^{-i k r} \mathrm{d} x_0 \mathrm{d} y_0
\end{equation}
$U_0$ is the field in the plane of the mask, $z$ is the distance from the mask to the plane where $U_1$ is being calculated, and $r$ is the distance from the point $(x_0,y_0)$ to the point $(x_1,y_1)$.  We can take the binomial expansion of the exponent, keeping only the first two terms:
\begin{align}
r & =  \sqrt{z^2+(x_0-x_1)^2+(y_0-y_1)^2} \nonumber \\
  & \approx z\left[1+\frac{1}{2}\left(\frac{x_0-x_1}{z}\right)^2 +\frac{1}{2}\left(\frac{y_0-y_1}{z}\right)^2 \right] \nonumber
\end{align}
Equation~\ref{field} becomes:
\begin{multline}
\label{field2}
U_1(x_1,y_1) = \int_{-\infty}^{\infty} \int_{-\infty}^{\infty} U_0(x_0,y_0) \frac{-z}{i \lambda r^2} \mathrm{e}^{-i k z} \\
 \mathrm{e}^{-\frac{i k}{2 z}\left[ (x_0 - x_1)^2 + (y_0 - y_1)^2 \right]} \mathrm{d} x_0 \mathrm{d} y_0 
\end{multline}
This is a form that will be useful in analyzing the exponent.

If we take the exponential term of Equation~\ref{field2} and multiply out the exponent, we get:
\begin{multline}
\mathrm{e}^{-\frac{i k}{2 z}\left[ (x_0 - x_1)^2 + (y_0 - y_1)^2 \right]} = \\
\mathrm{e}^{-\frac{i k}{2 z}\left[ x_0^2 + y_0^2 \right]} \mathrm{e}^{-\frac{i k}{2 z}\left[ x_1^2 + y_1^2 \right]} \mathrm{e}^{- \frac{i k}{2 z}\left[ x_0 x_1 + y_0 y_1 \right]} \nonumber
\end{multline}

For an exact Fourier transform, we only want to have the $\mathrm{e}^{-\frac{i k}{z}\left[ x_0 x_1 + y_0 y_1 \right]}$ term inside the integral.  The $\mathrm{e}^{-\frac{i k}{2 z}\left[ x_1^2 + y_1^2 \right]}$ term can go outside the integral, but we still have to deal with $\mathrm{e}^{-\frac{i k}{2 z}\left[ x_0^2 + y_0^2 \right]}$. Putting a lens behind the mask results in a phase shift given by:
\begin{equation}
A(x_0,y_0) = \mathrm{e}^{-i k n \Delta} \mathrm{e}^{\frac{i k}{2 z}\left[ x_0^2 + y_0^2 \right]} \nonumber
\end{equation}
$f$ is the focal length of the lens.  This exactly cancels $\mathrm{e}^{-\frac{i k}{z}\left[ x_0^2 + y_0^2 \right]}$ if $z = f$.  The resulting field is given by:
\begin{multline}
U_1(x_1,y_1) = \mathrm{e}^{-\frac{i k}{2 z}\left[ x_1^2 + y_1^2 \right]} \mathrm{e}^{-i k n \Delta} \frac{-z}{i \lambda r^2} \\
\int_{-\infty}^{\infty} \int_{-\infty}^{\infty} U_0(x_0,y_0) \mathrm{e}^{-\frac{i k}{z}\left[ x_0 x_1 + y_0 y_1 \right]} \mathrm{d} x_0 \mathrm{d} y_0
\end{multline}
So, the focal plane of a lens behind a mask is an exact Fourier transform of the plane containing the mask.

The inverse transform can be performed in the same manner as the initial transform.  This is due to the symmetry of the Fourier transform plane: $U_0(x_0,y_0) = U_0(-x_0,-y_0)$, allowing us to reverse the sign in the exponent.  However, there is still the factor of $\mathrm{e}^{-\frac{i k}{z}\left[ x_0 x_1 + y_0 y_1 \right]}$ outside the integral that will create problems.  Our next lens must cancel both $\mathrm{e}^{-\frac{i k}{z_1}\left[ x_0 x_1 + y_0 y_1 \right]}$ and $\mathrm{e}^{-\frac{i k}{z_2}\left[ x_0 x_1 + y_0 y_1 \right]}$.  To do this, we must satisfy the equations:
\begin{equation}
\label{inexact_ft}
\frac{1}{z_1} - \frac{1}{z_2} = - \frac{1}{f_2}
\end{equation}
Point-wise multiplying the field values after the first lens cause the term outside the integral to be squared: $\mathrm{e}^{\frac{i k}{z}\left(x_1^2+y_1^2\right)}$.  To achieve an exact inverse Fourier transform, we must therefore satisfy the equation:
\begin{equation}
\label{exact_ft}
\frac{2}{z_1} - \frac{1}{z_2} = - \frac{1}{f_2}
\end{equation}
In general, the $i^{th}$ lens must be at a distance: 
\begin{equation}
\label{gen_ft}
z_i = - \frac{1}{\frac{1}{f_i}-\frac{n+1}{z_{i-1}}}
\end{equation}
$n$ is the number of multiplies done between steps $i$ and $i-1$.
\begin{figure}[h]
\centering
\includegraphics[width=0.4 \textwidth]{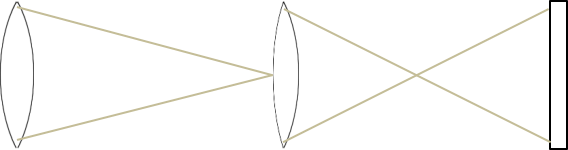}
\caption{Lens configuration to produce an exact Fourier transform}
\end{figure}

\section{Cropping}
An experiment was done by simulation to explore the effect of different parameters on how much information is lost when a Fourier transform image is cropped.  In this experiment, we put light in a checkerboard pattern through two lenses a distance $2f$ apart and put the output screen a distance $2f$ from the second lens.  Though this is not an exact Fourier transform, it nevertheless returns the original image.  
\begin{figure}[h]
\centering
\includegraphics[width=0.4 \textwidth]{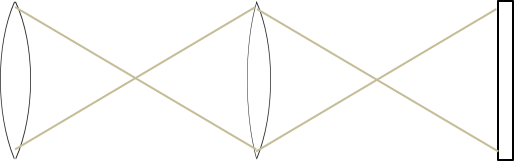}
\caption{Lens configuration to produce an inexact Fourier transform}
\end{figure}

We ran the experiment a number of times with different values of wavelength and focal length and compared the output images.  The field at the second lens was cropped to different sizes.  As expected, larger cropping sizes yielded a much clearer output image than the smaller ones.  As wavelength decreased, the output image also became clearer.  This verifies the expectation that a shorter wavelength results in a more compact Fourier transform.  The output image improved for a shorter focal length, showing that a large focal length contributes to the spreading out of a Fourier transform.  These results are shown in the table below.
\begin{figure}[h]
\centering
\begin{tabular}{|c|c|}
\hline
\includegraphics[width=0.16 \textwidth]{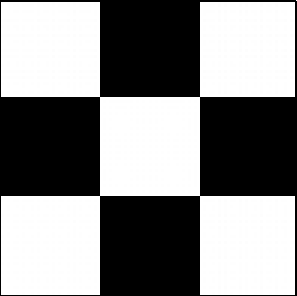} & \includegraphics[width=0.16 \textwidth]{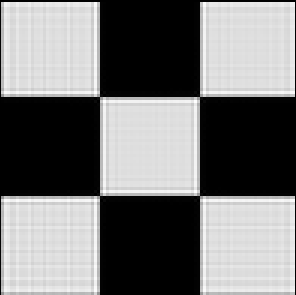} \\
{\footnotesize $x = 4 \, \mathrm{mm}$} & {\footnotesize $x = 3 \, \mathrm{mm}$} \\
\hline
\includegraphics[width=0.16 \textwidth]{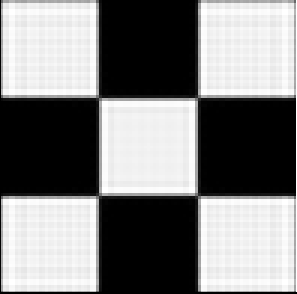} & \includegraphics[width=0.16 \textwidth]{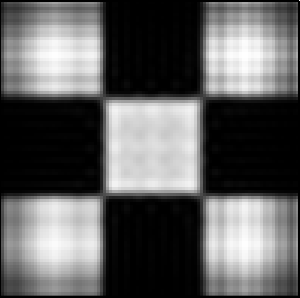} \\
{\footnotesize $x = 2 \, \mathrm{mm}$} & {\footnotesize $x = 1 \, \mathrm{mm}$} \\
\hline
\end{tabular}
\caption{Effects of changing cropping size ($\lambda = 200 \, \mathrm{nm}$, $f = 100 \, \mathrm{mm}$)}
\end{figure}
\begin{figure}[h]
\centering
\begin{tabular}{|c|c|}
\hline
\includegraphics[width=0.16 \textwidth]{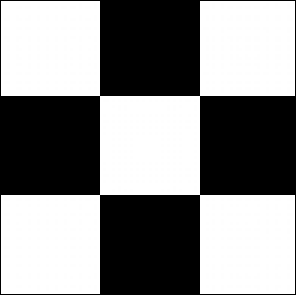} & \includegraphics[width=0.16 \textwidth]{200nm_100mm_2x2mm_c.PNG} \\
{\footnotesize $\lambda = 100 \, \mathrm{nm}$} & {\footnotesize $\lambda = 200 \, \mathrm{nm}$} \\
\hline
\includegraphics[width=0.16 \textwidth]{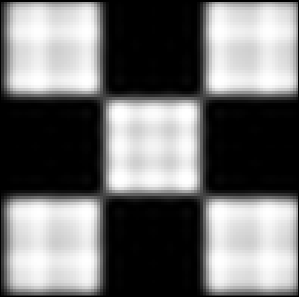} & \includegraphics[width=0.16 \textwidth]{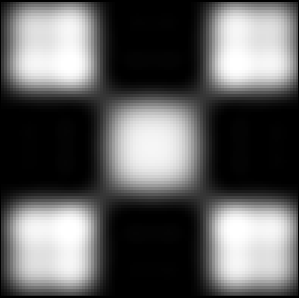} \\
{\footnotesize $\lambda = 600 \, \mathrm{nm}$} & {\footnotesize $\lambda = 1200 \, \mathrm{nm}$} \\
\hline
\end{tabular}
\caption{Effects of changing wavelength ($x = 2 \, \mathrm{mm}$, $f = 100 \, \mathrm{mm}$)}
\end{figure}
\begin{figure}[h]
\centering
\begin{tabular}{|c|c|}
\hline
\includegraphics[width=0.16 \textwidth]{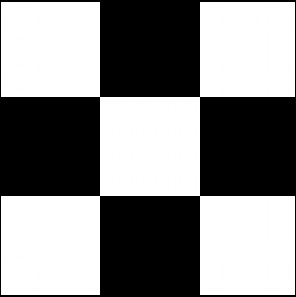} & \includegraphics[width=0.16 \textwidth]{200nm_100mm_2x2mm_c.PNG} \\
{\footnotesize $f = 50 \, \mathrm{mm}$} & {\footnotesize $f = 100 \, \mathrm{mm}$} \\
\hline
\includegraphics[width=0.16 \textwidth]{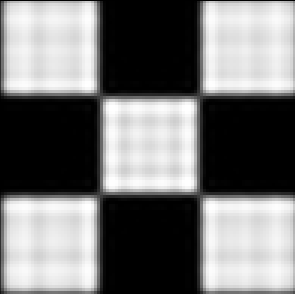} & \includegraphics[width=0.16 \textwidth]{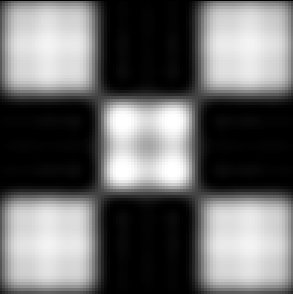} \\
{\footnotesize $f = 200 \, \mathrm{mm}$} & {\footnotesize $f = 400 \, \mathrm{mm}$} \\
\hline
\end{tabular}
\caption{Effects of changing focal length ($x = 2 \, \mathrm{mm}$, $\lambda = 200 \, \mathrm{nm}$)}
\end{figure}

\section{Point-wise Multiplication}
Performing a point-wise multiply optically remains a challenge.  We must encode the digits of our numbers spatially in the amplitudes of the input field in order to obtain the Fourier transform.  However, combining information contained in the amplitudes of two fields results in a sum offset by phase differences, not a product.  A product could be obtained by passing light from one transformed input through an amplitude mask representing the other transformed input.  For example, if the amplitude at a specific point on one of the inputs was 30\% of the peak amplitude, it would correspond to a point on the mask that let 30\% of the light through.  If the corresponding point in the other input was 70\% of the peak amplitude, after going through the mask it would be 21\% of the peak, effectively representing $3 \times 7$.  The problem with this is that we would have to make a new mask for every pair of inputs.  It is feasible to use some photochromic material, exposed to the diffraction pattern of a light source, to create such a mask during computation.  However, this process is likely to be slow and would create a bottleneck in computation.

If our goal is to raise a number to a power, it may be worthwhile to make a mask and then pass the input through many times.  However, without the ability to do carries, the number of distinct integers stored in each point would quickly grow to exceed our ability to distinguish with a detector.  Physically, passing the same ray of light through a mask many times will cause it to grow fainter and fainter until the signal is lost.

Using phase differences to compute multiplication is another option.  However, it would be a difficult task of itself to calculate how much to shift the phase in order to obtain the correct combined value.  We need to solve the equation:
\begin{gather}
A \mathrm{e}^{-i k t} + B \mathrm{e}^{-i (k t + \varphi)} = A B \mathrm{e}^{-i k t} \nonumber \\
B \mathrm{e}^{-i (k t + \varphi)} = (A B - A) \mathrm{e}^{-i k t} \nonumber \\
\mathrm{e}^{-i (k t + \varphi)} = \frac{A}{B}(B-1) \mathrm{e}^{-i k t} \nonumber \\
-i (k t + \varphi)= \ln\left[\frac{A}{B}(B-1) \mathrm{e}^{-i k t}\right] \nonumber \\
\varphi = i \ln\left[\frac{A}{B}(B-1)\right] \nonumber
\end{gather}
This involves a fair amount of computation that would need to be handled by an auxiliary computer, so it is easier to consider simply doing the point-wise multiply with a detector and electronic computer.  The result of the multiply would then be encoded in light to be put through another lens to do the inverse Fourier transform.  Phase information is essential for correctly performing an inverse Fourier transform, so the output beam would need to incorporate phase.

\section{Encoding Digits}
There are several options for how to encode numbers into light patterns.  The most obvious way is to set equally sized bands to different amplitudes.  This works reasonably well, and the results of several multiplication problems are shown below.  Note that the output plane is the mirror image of the answer, so digits must be read from right to left.
\begin{figure}[h]
\centering
\begin{tabular}{|c|c|c|}
\hline
\includegraphics[width=0.13 \textwidth]{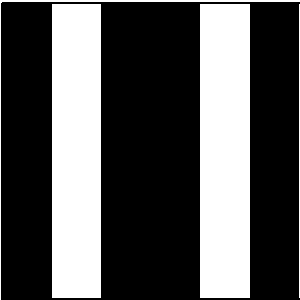} & \includegraphics[width=0.13 \textwidth]{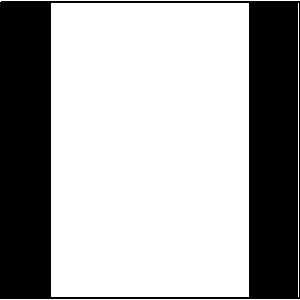} & \includegraphics[width=0.13 \textwidth]{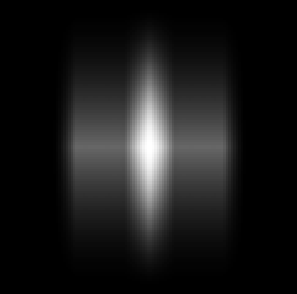} \\
{\tiny 010010} & {\tiny 011110} & {\tiny 0011121110} \\
\hline
\includegraphics[width=0.13 \textwidth]{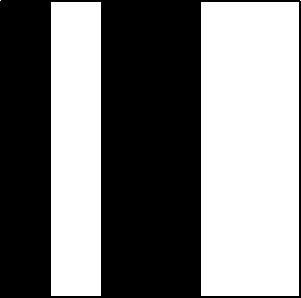} & \includegraphics[width=0.13 \textwidth]{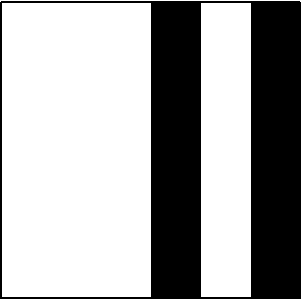} & \includegraphics[width=0.13 \textwidth]{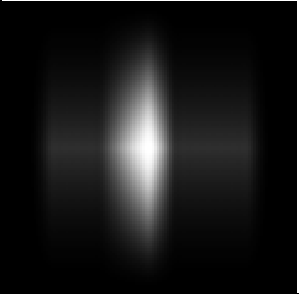} \\
{\tiny 010011} & {\tiny 111010} & {\tiny 01111321110} \\
\hline
\includegraphics[width=0.13 \textwidth]{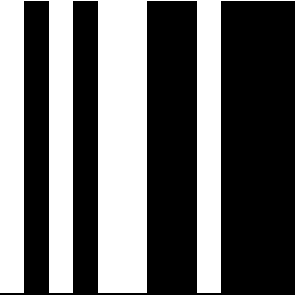} & \includegraphics[width=0.13 \textwidth]{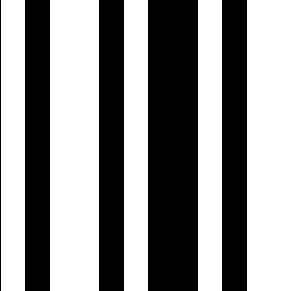} & \includegraphics[width=0.13 \textwidth]{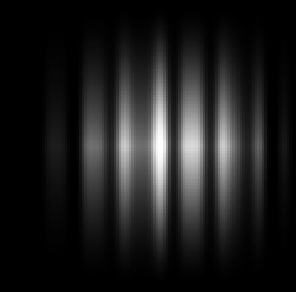} \\
{\tiny 101011001000} & {\tiny 101101001011} & {\tiny 010212313313223122011000} \\
\hline
\end{tabular}
\caption{Results of multiplication using banded encoding}
\end{figure}
\begin{figure}[h]
\centering
\begin{tabular}{|c|c|c|}
\hline
\includegraphics[width=0.13 \textwidth]{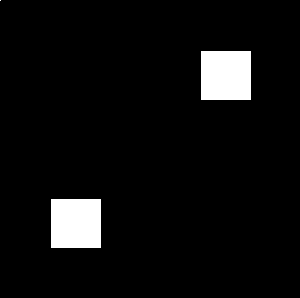} & \includegraphics[width=0.13 \textwidth]{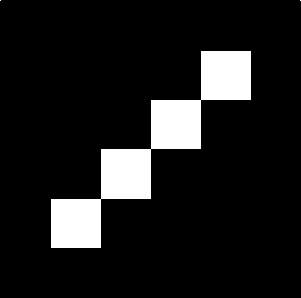} & \includegraphics[width=0.13 \textwidth]{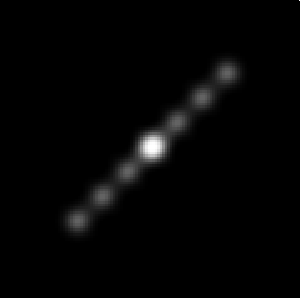} \\
{\tiny 010010} & {\tiny 011110} & {\tiny 0011121110} \\
\hline
\includegraphics[width=0.13 \textwidth]{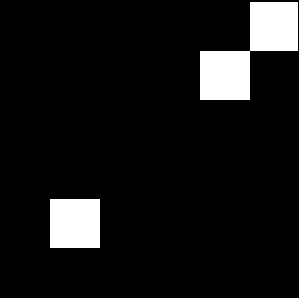} & \includegraphics[width=0.13 \textwidth]{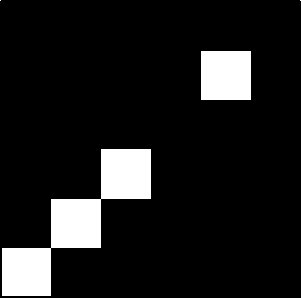} & \includegraphics[width=0.13 \textwidth]{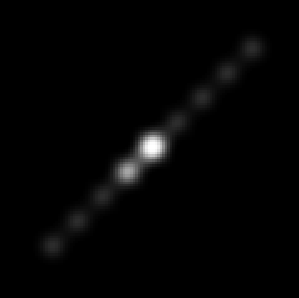} \\
{\tiny 010011} & {\tiny 111010} & {\tiny 01111321110} \\
\hline
\includegraphics[width=0.13 \textwidth]{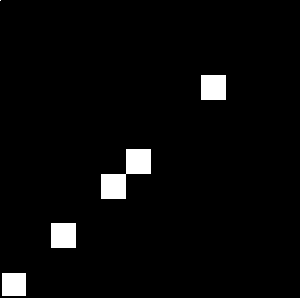} & \includegraphics[width=0.13 \textwidth]{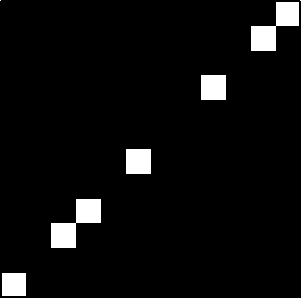} & \includegraphics[width=0.13 \textwidth]{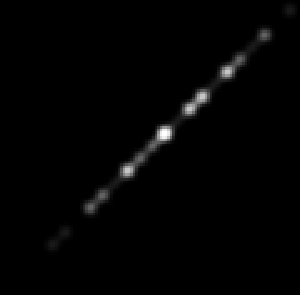} \\
{\tiny 101011001000} & {\tiny 101101001011} & {\tiny 010212313313223122011000} \\
\hline
\end{tabular}
\caption{Results of multiplication using diagonal encoding}
\end{figure}
It is possible to see the digits in the bands of the output, but it is rather difficult to tell where each digit starts and ends.  The picture becomes much clearer when we utilize both dimensions of the plane and encode our digits diagonally. For the remainder of this paper, we will encode binary digits numerically. When we apply this technique to the modular multiplication problem below, we will add additional spacing between digits to prevent adjacent digits from blurring together.

\section{Simulation Code}
Much good work has already been done in creating simulators for Fourier optics~\cite{louri94, trester99, rafferty10}. Our simulator consists of approximately 1800 lines of C code compiled on a Linux workstation using gcc. It is parallelized using the OpenMP library to distribute the 2D convolution workload over threads. Profiling with gprof demonstrates that the code spends at least 95\% of its time computing the Fresnel approximation for lenses. Runtimes vary with the pixel count of the image. Running on 20 threads of a dual socket Intel(R) Xeon(R) CPU E5-2650 v3 @ 2.30GHz node, the simulation time ranges from around 25 seconds per lens at $250 \times 250$ pixels resolution to 80 minutes per lens at $1000 \times 1000$ pixels resolution.

The simulation is configured using an instruction file lists the devices to be modeled during each step of the execution. An example instruction file is shown in Figure~\ref{instr_file} below. The code can maintain arbitrarily many optical paths that can be split and recombined over the course of the simulation. The instruction file also includes the ability to specify ``tap'' files to write out intermediate data values. The simulation results were validated by comparison with rectangular aperture results documented by Abedin~\cite{abedin07}.

\section{Modular Multiplication}
Modular multiplication involves solving problems of the form $c = a \cdot b \!\! \mod m$ for integer values of $a$, $b$ and $m$. Typically, the value of each of these will be of similar magnitude. What makes solving this problem difficult on modern computers is the fact that the modulo, or remainder, operation requires a division which is an expensive operation. However, for cases in which the modulus $m$ is odd, Montgomery has proposed an alternate formulation of the problem that replaces division by $m$ with division by an alternate base $r$ that is chosen to be a power of two greater than $m$. For binary arithmetic, this change reduces the costly division problem to one of inexpensive masks and shifts~\cite{montgomery85, warren}.

\subsection{Montgomery multiplication}
Choose a value $r = 2^k$ to be the next power of two greater than $m$ and compute $M = -m^{-1} \!\! \mod r$. Montgomery observed that when the values of $a$ and $b$ are transformed to new values $\bar{a} = a \cdot r \!\! \mod m$ and $\bar{b} = b \cdot r \!\! \mod m$ then we can calculate the corresponding value of $\bar{c}$ as follows:
\begin{equation}
\bar{c} = \frac{\bar{a}\bar{b} + \left(\bar{a}\bar{b}M \!\!\!\!\! \mod r\right)m}{r}
\label{Montgomery}
\end{equation}
We observe that for an arbitrary integer $x$, the value of $x \!\! \mod r$ is the lower $k$ bits of the binary representation of $x$ while division by $r$ is the upper $|x|-r$ bits. To recover the solution $c$ one needs to convert the result back out of the Montgomery domain which is accomplished by setting $c = \bar{c} \cdot R \!\! \mod m$ where $R = r^{-1} \!\! \mod m$. The example below illustrates the Montgomery multiplication on 16-bit values of $a$, $b$ and $m$.

\subsection{Montgomery example}
Compute $a \cdot b \!\! \mod m$ using $a = 28510$, $b = 38762$ and $m = 36057$. Seeing that $a$, $b$ and $m$ are all 16-bit values, choose $r = 2^{17} = 131072$ and pre-compute $M$ and $R$.
\begin{align}
M & = - 36057^{-1} \!\!\!\! \mod 131072 = 52375 \nonumber \\
R & = 131072^{-1} \!\!\!\! \mod 36057 = 14408 \nonumber
\end{align}
Next, convert $a$ and $b$ into the Montgomery basis.
\begin{align}
\bar{a} & = 28510 \cdot 131072 \!\!\!\! \mod 36057 = 23411 \nonumber \\
\bar{b} & = 38762 \cdot 131072 \!\!\!\! \mod 36057 = 31495 \nonumber
\end{align}
Now, the steps of the Montgomery multiplication proceed as shown below.
\begin{align}
k_1 & = \bar{a}\bar{b} \nonumber \\
       & = 23411 \cdot 31495 = 737329445 \nonumber \\
k_2 & = k_1 \cdot M \nonumber \\
       & = 737329445 \cdot 52375 = 38617629681875 \nonumber \\
k_3 & = k_2 \!\!\!\! \mod r \nonumber \\
       & = \mbox{AND} \! \left( 38617629681875, 131071 \right) = 92371 \nonumber \\
k_4 & = k_3 \cdot m \nonumber \\
       & = 92371 \cdot 36057 = 3330621147 \nonumber \\
k_5 & = k_1 + k_4 \nonumber \\
       & = 737329445 + 3330621147 = 4067950592 \nonumber \\
\bar{c} & = k_5 / r \nonumber \\
       & = \mbox{SHIFT} \! \left( 4067950592, 17 \right) = 31036 \nonumber
\end{align}
Convert the result back out of the Montgomery domain.
\begin{equation}
c = \bar{c} \cdot R \!\!\!\! \mod m = 31036 \cdot 14408 \!\!\!\! \mod 36057 = 23831 \nonumber
\end{equation}
we see that this is in fact the correct solution to our initial problem: $28510 \cdot 38762 \!\! \mod 36057 = 23831$. Looking back at the calculation above, there are a few important simplifications which will make the problem easier to implement in optics. They will be discussed in the next section.

\subsection{Distinguishing high and low order bits}
{\color{black} As described by Warren~\cite{warren}, a} closer inspection of Equation~\ref{Montgomery} reveals that the lower $k$ bits of the numerator are not required, since they will be truncated in the division by $r$. Thus it is sufficient to compute the high bits of each of the values in the numerator, with two additional bits to address carry. Additionally, the $\mod r$ in the numerator will eliminate the high bits of $\bar{a}\bar{b} M$, so they do not need to be calculated either. Taking these observations into account, the example above now proceeds as follows.
\begin{align}
k_1 & = \bar{a}\bar{b} \nonumber \\
       & = 23411 \cdot 31495 = 737329445 \nonumber \\
k_{1_{lo}} & = \mbox{AND} \! \left( 737329445, 131071 \right) = 49455 \nonumber \\
k_{1_{hi}} & = \mbox{RSHIFT} \! \left( 737329445, 15 \right) = 22501 \nonumber \\
k_2 & = k_{1_{lo}} \cdot M \nonumber \\
       & = 49455 \cdot 52375 = 2589681875 \nonumber \\
k_3 & = k_2 \!\!\!\! \mod r \nonumber \\
       & = \mbox{AND} \! \left( 2589681875, 131071 \right) = 92371 \nonumber \\
k_4 & = k_3 \cdot m \nonumber \\
       & = 92371 \cdot 36057 = 3330621147 \nonumber \\
k_{4_{hi}} & = \mbox{SHIFT} \! \left( 3330621147, 15 \right ) = 101642 \nonumber \\
k_{5_{hi}} & = k_{1_{hi}} + k_{4_{hi}} + 1 \nonumber \\
       & = 22501 + 101642 + 1 = 124144 \nonumber \\
\bar{c} & = k_{5_{hi}} / 2^2 \nonumber \\
       & = \mbox{SHIFT} \! \left( 124144, 2 \right) = 31036 \nonumber
\end{align}
Notice that $k_{1_{hi}}$ and $k_{4_{hi}}$ are being shifted by only 15 bits, which is two fewer bits than the size of $r$. Those two overlap bit prevent carry errors from propagating from the less significant bits during addition. However, those two additional bits still need to be truncated in the final step when computing $\bar{c}$. Also, note that when computing $k_{5_{hi}}$ there is an extra one being added. This serves the purpose of setting the carry bit resulting from the addition of the less significant bits. {\color{black} Where this technique differs from the standard approach described by~\cite{warren} is in the treatment of the addition to compute $k_5$ from the example above. Warren describes the addition as involving both upper and lower bits of $k_1$, while we have discarded all but a few ``overlap'' digits by the time we reach this step. This will be significant in optical implementation, described in the next section, because we must demonstrate the ability to perform this approximation not for a multiplication but for a convolution instead.}

\subsection{Approximate implementation}
The simplifications of the previous section will form the basis of an optical approach to computing the modular multiplication. Limiting the number of bits carried from one operation to the next reduce the optical resolution required to accurately compute a solution, thus enabling the computation of larger numbers of digits. The intuition behind this is that the frequency of the Fourier transform will scale with the number of digits in the optically encoded representation of each number, and the resolution required to accurately evaluate the point-wise multiplication will be bounded by the Nyquist frequency and the size of the image in the transform plane.

\begin{figure}[h]
\centering
\includegraphics[width=0.45 \textwidth]{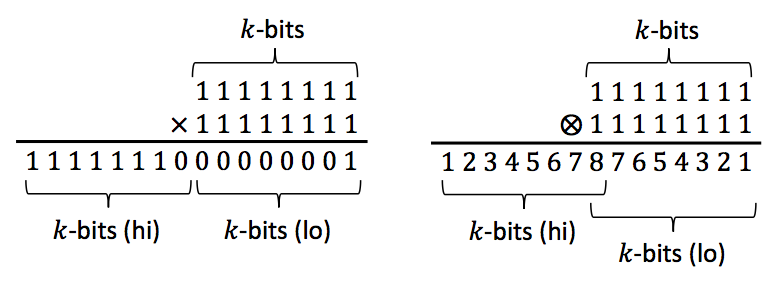}
\caption{Comparing multiplication on the left and convolution on the right in terms of the high and low $k$-digits of the result. When masking high order bits of the convolution operation, additional bits are necessary to capture the unpropagated carries from the low bits. Futher, the middle bit is counted as both a high and low bit, which will result in the shift of one additional digits when computing our approximate $\bar{c}$ below.}
\label{mult_vs_conv}
\end{figure}

Also note that there are some important distinctions between Montgomery in its digital implementation and the new hybrid optical approach. The most obvious difference is the fact that ``multiplications'' in optics are not actually multiplications but convolutions, as illustrated in Figure~\ref{mult_vs_conv}. Convolution requires one fewer digit in its representation than multiplication. However, the value of the digits grow beyond the magnitude of the base because the convolution performs no carries from one digit to the next. {\color{black} The consequence is that the two ``bits'' of overlap required previously to prevent carry errors from propagating will now translate to a number of ``digits'' of overlap. Since those digits  contain unresolved carries, more digits will be needed, but this is a complicated calculation and no formal upper bound has yet been established. For the purposes of the example in this paper we performed experiments to determine by trial-and-error that {\em six} digts would be a sufficient to prevent carry errors.}

In terms of the optical computation, this approach trades off resolution for increased dynamic range. It must also contend with the fact that shifting and masking does not translate directly to the results of convolution arithmetic. For clarity, values are written as a vector of their binary digits.
\begin{align}
 \bar{a} & = (1,\!0,\!1,\!1,\!0,\!1,\!1,\!0,\!1,\!1,\!1,\!0,\!0,\!1,\!1) \nonumber \\
 \bar{b} & = (1,\!1,\!1,\!1,\!0,\!1,\!1,\!0,\!0,\!0,\!0,\!0,\!1,\!1,\!1) \nonumber \\
 m & = (1,\!0,\!0,\!0,\!1,\!1,\!0,\!0,\!1,\!1,\!0,\!1,\!1,\!0,\!0,\!1) \nonumber \\
 M &= (1,\!1,\!0,\!0,\!1,\!1,\!0,\!0,\!1,\!0,\!0,\!1,\!0,\!1,\!1,\!1) \nonumber
 \end{align}

Now  reproduce the steps from the preceding section as follows, using {\em six} overlap digits instead of two. The notation $x \otimes y$ will denote  the digit convolution of $x$ and $y$, and $x \oplus y$ will denote the digit sum.
\begin{align}
k_1 = & \; \bar{a} \otimes \bar{b} \nonumber \\
       = & \; (1,\!1,\!2,\!3,\!2,\!4,\!4,\!3,\!5,\!4,\!4,\!5,\!4,\!4,\!6,\!6,\!5,\!3,\!3, \nonumber \\
       & \; \; 4,\!3,\!2,\!3,\!2,\!1,\!1,\!2,\!2,\!1) \nonumber \\
       = & \; 737329445 \nonumber \\
k_{1_{hi}} = & \; (1,\!1,\!2,\!3,\!2,\!4,\!4,\!3,\!5,\!4,\!4,\!5,\!4,\!4,\!6,\!6,\!5,\!3,\!3) \nonumber \\
       = & \; 720045 \nonumber \\
k_{1_{lo}} = & \; (4,\!4,\!6,\!6,\!5,\!3,\!3,\!4,\!3,\!2,\!3,\!2,\!1,\!1,\!2,\!2,\!1) \nonumber \\
       = & \; 573733 \nonumber \\
k_2 = & \; k_{1_{lo}} \otimes M \nonumber \\
       = & \; (4,\!8,\!10,\!12,\!15,\!16,\!16,\!19,\!22,\!17,\!17,\!22, \nonumber \\
       & \; \; 19,\!20,\!25,\!32,\!28,\!25,\!24,\!20,\!16,\!15,\!13, \nonumber \\
       & \; \; 11,\!9,\!8,\!6,\!5,\!5,\!5,\!3,\!1) \nonumber \\
       = & \; 30049265875 \nonumber \\
k_3 = & \;  k_2 \!\!\!\! \mod r \nonumber \\
       = & \; (32,\!28,\!25,\!24,\!20,\!16,\!15,\!13,\!11,\!9,\!8,\!6,\!5, \nonumber \\
       & \; \; 5,\!5,\!3,\!1) \nonumber \\
       = & \; 3762387 \nonumber \\
k_4 = & \; k_3 \otimes m \nonumber \\
       = & \; (32,\!28,\!25,\!24,\!52,\!76,\!68,\!62,\!87,\!105,\!92,\nonumber \\
       & \; \; 115,\!133,\!114,\!102,\!121,\!100,\!86,\!79,\!66, \nonumber \\
       & \; \; 51,\!43,\!37,\!30,\!23,\!19,\!14,\!9,\!6,\!5,\!3,\!1) \nonumber \\
       = & \; 135660388059 \nonumber \\
k_{4_{hi}} = & \; (32,\!28,\!25,\!24,\!52,\!76,\!68,\!62,\!87,\!105,\!92,\!115, \nonumber \\
       & \; \; 133,\!114,\!102,\!121,\!100,\!86,\!79,\!66,\!51,\!43) \nonumber \\
       = & \; 132480817 \nonumber \\
k_{5_{hi}} = & \; k_{1_{hi}} \oplus k_{4_{hi}} \oplus 1 \nonumber \\
       = & \; (32,\!28,\!25,\!25,\!53,\!78,\!71,\!64,\!91,\!109,\!95,\!120, \nonumber \\
       & \; \; 137,\!118,\!107,\!126,\!104,\!92,\!85,\!71,\!54,\!46) \nonumber \\
       = & \; 133200926 \nonumber \\
\bar{c} = & \; \left\lfloor k_{5_{hi}} / 2^7 \right\rfloor \nonumber \\
       = & \; \left\lfloor \left( \frac{1}{4},\frac{7}{32},\frac{25}{128},\frac{25}{128},\frac{53}{128},\frac{39}{64},\frac{71}{128},\frac{1}{2}, \right. \right. \nonumber \\
       & \; \; \frac{91}{128},\frac{109}{128},\frac{95}{128},\frac{15}{16},\frac{137}{128},\frac{59}{64},\frac{107}{128},\frac{63}{64}, \nonumber \\
       & \left. \left. \; \; \frac{13}{16},\frac{23}{32},\frac{85}{128},\frac{71}{128},\frac{27}{64},\frac{23}{64} \right) \right\rfloor \nonumber \\
       = & \; 1040632 \nonumber
\end{align}
This is a different value of $\bar{c}$ than obtained in the previous example. However, converting the result back out of the Montgomery domain with
\begin{equation}
c = \bar{c} \cdot R \!\!\!\! \mod m = 1040632 \cdot 14408 \!\!\!\! \mod 36057 = 23831 \nonumber
\end{equation}
demonstrates that it has recovered the correct value of $c$. The convolutional approach is an approximate modular multiplication in the sense that it produces a result that will typically be an arbitrary multiple of the modulus. Notice that in the last step of computing $\bar{c}$ the last \emph{seven} digits have been removed, despite the fact that the overlap was only six digits. That is because of the one fewer digits resulting from convolution compared to multiplication. The most significant $k$ digits of the result include one digits from the $k$ least significant digits which must also be removed.

\subsection{An optical implementation}
The computation of $28510 \cdot 38762 \!\! \mod 36057 = 23831$ was simulated in optics using the simulator and convolutional algorithmic approach described above. Shifts and masks are accomplished through optical masking and alignment. Addition is performed using a beam-splitter in reverse to combine values. Multiplications by one are performed to make certain that each value being added has passed through the same number of multiplier steps. This is necessary to keep the phase information consistent across the summands.
\begin{figure}[h]
\centering
\includegraphics[angle=90, width=0.4 \textwidth]{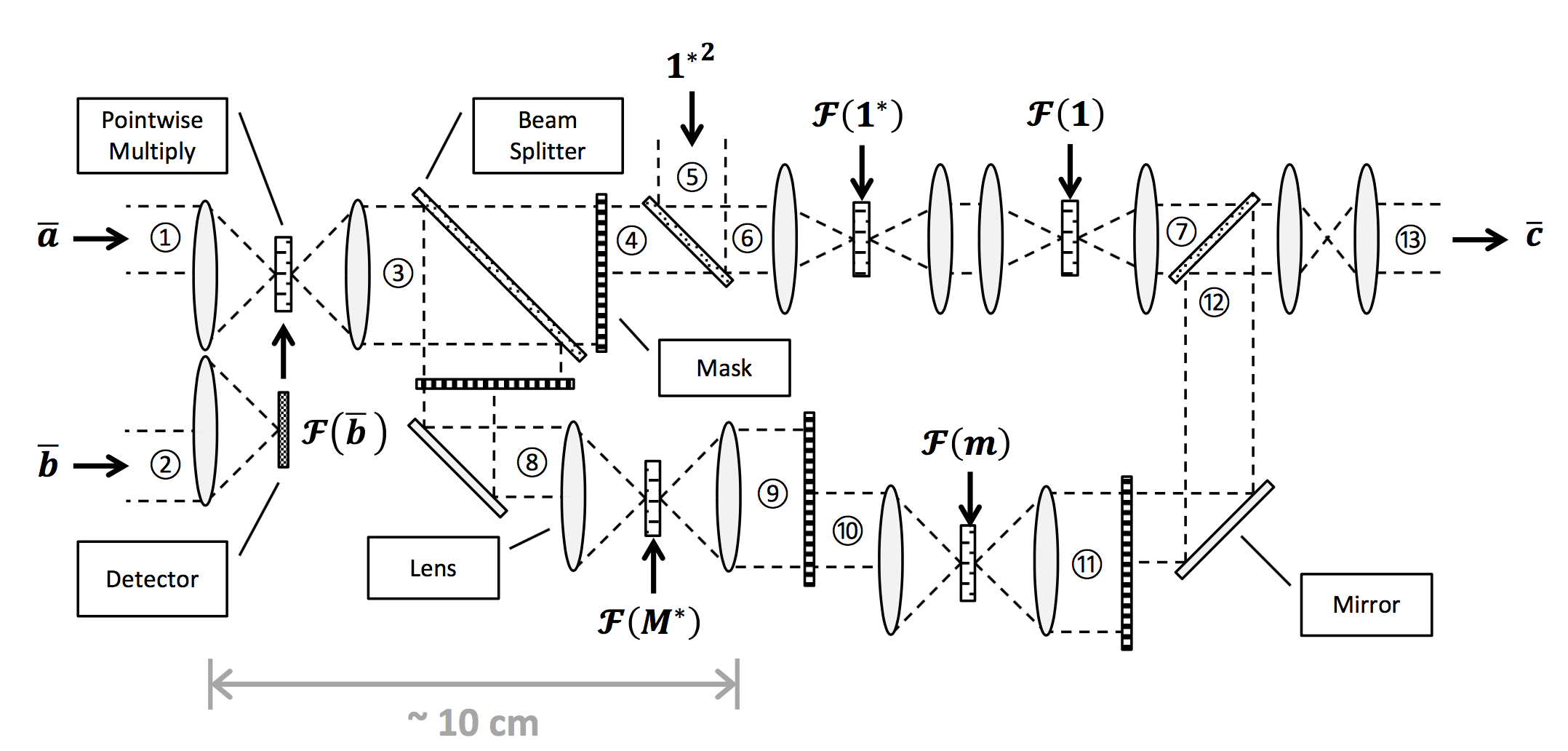}
\caption{The device configuration of a modular multiplication that receives inputs $\bar{a}$ and $\bar{b}$ in the Montgomery domain and produces an output $\bar{c}$, also in the Montgomery domain. Circled numbers represent locations where the image data is depicted in Figure~\ref{mod_mul_images} that follows.}
\label{lens_configuration}
\end{figure}

In general, the most significant bit (MSB) of a number is in the upper right hand corner of the image, and the least significant bit (LSB) is in the lower left hand corner. However, after each multiplication the resulting image is reversed. The reversal is denoted by an asterisk next to a value in the image. The optical configuration to implement the modular multiplication is illustrated in Figure~\ref{lens_configuration} above, and the instruction file to run the simulation is shown in Figure~\ref{instr_file} below. In general, the point-wise multiplies are assumed to be performed using a spatial light modulator (SLM) which creates an image mask of the Fourier transform of one of the values to be multiplied. As discussed above, this step is generally considered to be slow, but it is important to point out that only one of the SLMs, corresponding to the Fourier transform of $\bar{b}$ actually needs to be updated in real time. The remaining transforms can be precomputed, and thus the setting of the SLM once initialized will remain fixed. 
\begin{figure}[h]
\centering
{\tiny
\begin{verbatim}
    /home/jtdaly3/Software/opmul/mod_mul_16bit/output
    /home/jtdaly3/Software/opmul/mod_mul_16bit/tap
    0.000200
    1005
    #### Simulation: 28510 * 38672 mod 36057 = 23831
    generate 15 67 0 0 0 0 0 0 0 0 0 0 0 0 0 0 0 0 0 0 0 0 0 0 0 0 0 0 0 0 0 0
     0 0 0 0 1 0 1 0 0 0 0 0 1 0 1 0 1 0 0 0 1 0 1 0 0 0 1 0 1 0 0 0 1 0 0 0 0
    generate 15 67 0 0 1 0 1 0 1 0 0 0 0 0 0 0 0 0 0 0 1 0 1 0 0 0 1 0 1 0 1 0
     1 0 0 0 0 0 0 0 0 0 0 0 0 0 0 0 0 0 0 0 0 0 0 0 0 0 0 0 0 0 0 0 0 0 0 0 0
    generate 15 67 0 0 0 0 1 0 1 0 0 0 0 0 1 0 1 0 0 0 0 0 1 0 0 0 0 0 1 0 0 0
     1 0 1 0 1 0 0 0 0 0 0 0 0 0 0 0 0 0 0 0 0 0 0 0 0 0 0 0 0 0 0 0 0 0 0 0 0
    generate 15 67 0 0 0 0 0 0 0 0 0 0 0 0 0 0 0 0 0 0 0 0 0 0 0 0 0 0 0 0 0 0
     0 0 0 1 0 0 0 0 0 1 0 1 0 0 0 1 0 1 0 0 0 0 0 1 0 1 0 0 0 0 0 0 0 1 0 0 0
    generate 15 67 0 0 0 0 0 0 0 1 0 0 0 0 0 0 0 0 0 0 0 0 0 0 0 0 0 0 0 0 0 0
     0 0 0 0 0 0 0 0 0 0 0 0 0 0 0 0 0 0 0 0 0 0 0 0 0 0 0 0 0 0 0 0 0 0 0 0 0
    generate 15 67 0 0 0 0 0 0 0 0 0 0 0 0 0 0 0 0 0 0 0 0 0 0 0 0 0 0 0 0 0 0
     0 0 0 0 0 0 0 0 0 0 0 0 0 0 0 0 0 0 0 0 0 0 0 0 0 0 0 0 0 0 0 1 0 0 0 0 0
    generate 15 67 0 0 0 0 0 0 0 0 0 0 0 0 0 0 0 0 0 0 0 0 0 0 0 0 0 0 0 0 0 0
     0 0 0 0 0 0 0 0 0 0 0 0 0 0 0 0 0 0 0 0 1 0 0 0 0 0 0 0 0 0 0 0 0 0 0 0 0
    generate 15 67 0 0 0 0 0 0 0 0 0 0 0 0 0 0 0 0 0 0 0 0 0 0 0 0 0 0 0 0 0 0
     0 0 0 0 0 0 0 0 0 0 0 0 0 0 0 0 0 0 0 1 0 0 0 0 0 0 0 0 0 0 0 0 0 0 0 0 0
    tap
    lens 1 15.000000 -15.000000 1.500000 2.000000 15.000000 1.000000 1.000000
    lens 2 15.000000 -15.000000 1.500000 2.000000 15.000000 1.000000 1.000000
    tap
    pointwise_mul 1 2
    lens 1 5.000000 -5.000000 1.500000 2.000000 15.000000 1.000000 1.000000
    tap
    beam_splitter 1
    mask 1 -0.500000 0.164179 0.164179 -0.500000
    mask 2 -0.044776 0.500000 0.500000 -0.044776
    tap
    lens 2 7.500000 -7.500000 1.500000 2.000000 15.000000 1.000000 1.000000
    lens 3 15.000000 -15.000000 1.500000 2.000000 15.000000 1.000000 1.000000
    tap
    pointwise_mul 2 3
    lens 2 5.000000 -5.000000 1.500000 2.000000 15.000000 1.000000 1.000000
    tap
    mask 2 -0.500000 0.029851 0.029851 -0.500000
    tap
    lens 2 7.500000 -7.500000 1.500000 2.000000 15.000000 1.000000 1.000000
    lens 3 15.000000 -15.000000 1.500000 2.000000 15.000000 1.000000 1.000000
    tap
    pointwise_mul 2 3
    lens 2 5.000000 -5.000000 1.500000 2.000000 15.000000 1.000000 1.000000
    tap
    mask 2 -0.500000 0.179104 0.179104 -0.500000
    tap
    lens 3 15.000000 -15.000000 1.500000 2.000000 15.000000 1.000000 1.000000
    lens 4 15.000000 -15.000000 1.500000 2.000000 15.000000 1.000000 1.000000
    tap
    pointwise_mul 3 4
    lens 3 5.000000 -5.000000 1.500000 2.000000 15.000000 1.000000 1.000000
    tap
    filter 3 0.5
    tap
    pointwise_add 1 3
    tap
    lens 1 7.500000 -7.500000 1.500000 2.000000 15.000000 1.000000 1.000000
    lens 3 15.000000 -15.000000 1.500000 2.000000 15.000000 1.000000 1.000000
    tap
    pointwise_mul 1 3
    lens 1 5.000000 -5.000000 1.500000 2.000000 15.000000 1.000000 1.000000
    tap
    lens 1 7.500000 -7.500000 1.500000 2.000000 15.000000 1.000000 1.000000
    lens 3 15.000000 -15.000000 1.500000 2.000000 15.000000 1.000000 1.000000
    tap
    pointwise_mul 1 3
    lens 1 5.000000 -5.000000 1.500000 2.000000 15.000000 1.000000 1.000000
    tap
    pointwise_add 1 2
    tap
    lens 1 7.500000 -7.500000 1.500000 2.000000 15.000000 1.000000 1.000000
    lens 1 7.500000 -7.500000 1.500000 2.000000 15.000000 1.000000 1.000000
    tap
    read_out 1
    detector 1 12.000000 67
\end{verbatim}
}
\caption{Instruction file to simulate 16-bit modular multiplication illustrated in Figure~\ref{lens_configuration} at 1 $\!\mu$m pixel pitch and 15 $\!$mm separation between lenses.}
\label{instr_file}
\end{figure}

\begin{figure*}
\centering
\begin{tabular}{c}
\begin{tabular}{|c|c|c|c|}
\hline
\includegraphics[width=0.2 \textwidth]{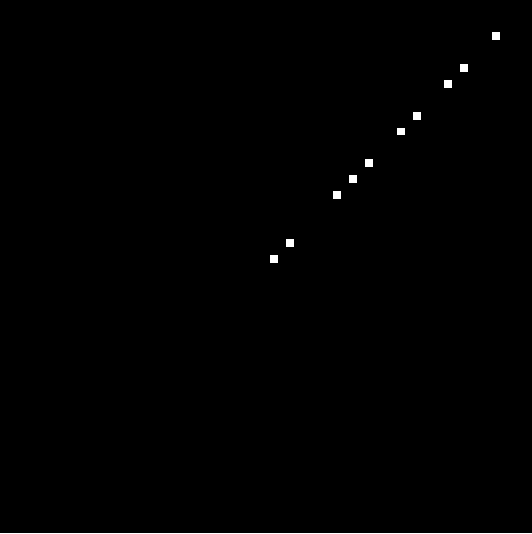} &
\includegraphics[width=0.2 \textwidth]{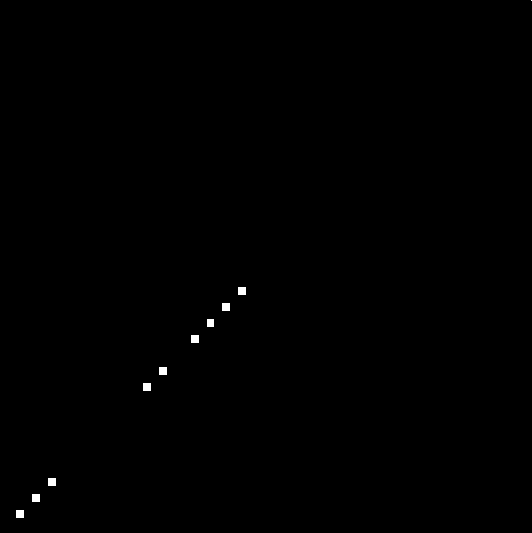} &
\includegraphics[width=0.2 \textwidth]{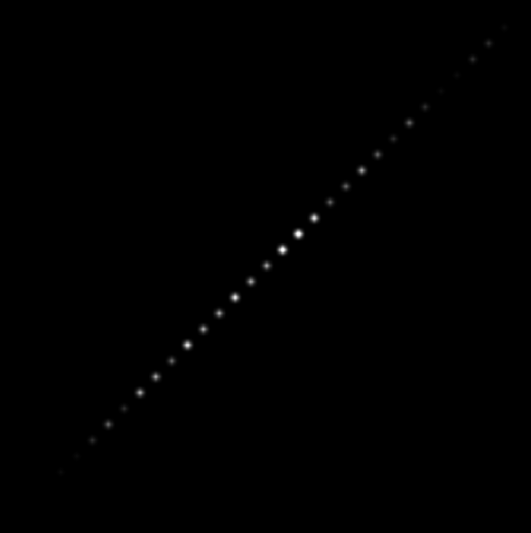} &
\includegraphics[width=0.2 \textwidth]{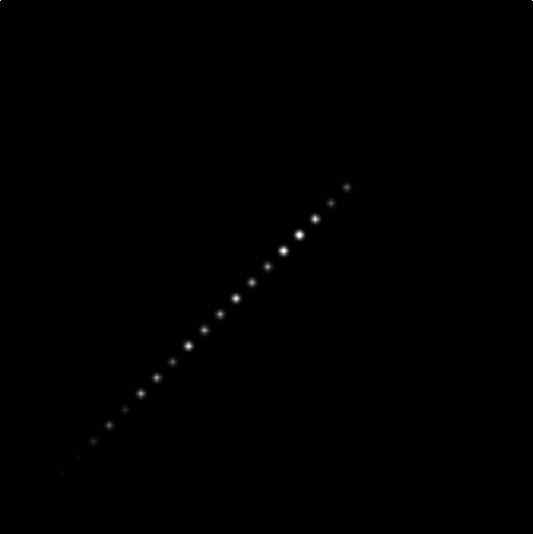} \\
{\tiny \#1 (1,0,1,1,0,1,1,0,1,1,1,0,0,1,1)} &
{\tiny \#2 (1,1,1,1,0,1,1,0,0,0,0,0,1,1,1)} &
{\tiny \#3 \begin{tabular}{c}(1,1,2,3,2,4,4,3,5,4,4,5,4,4,6, \\
6,5,3,3,4,3,2,3,2,1,1,2,2,1)*\end{tabular}}&
{\tiny \#4 (1,1,2,3,2,4,4,3,5,4,4,5,4,4,6,6,5,3,3)*} \\
\hline
\includegraphics[width=0.2 \textwidth]{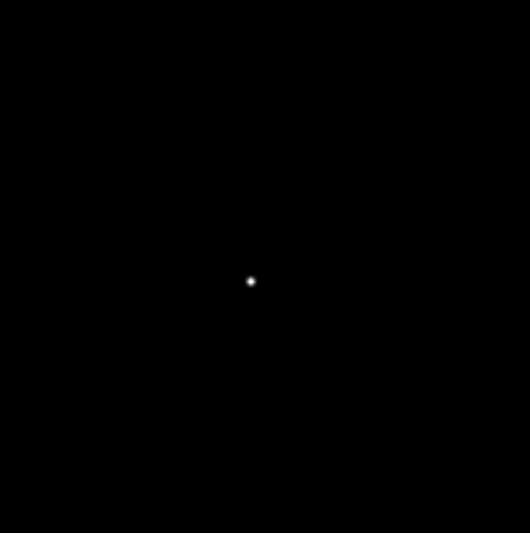} &
\includegraphics[width=0.2 \textwidth]{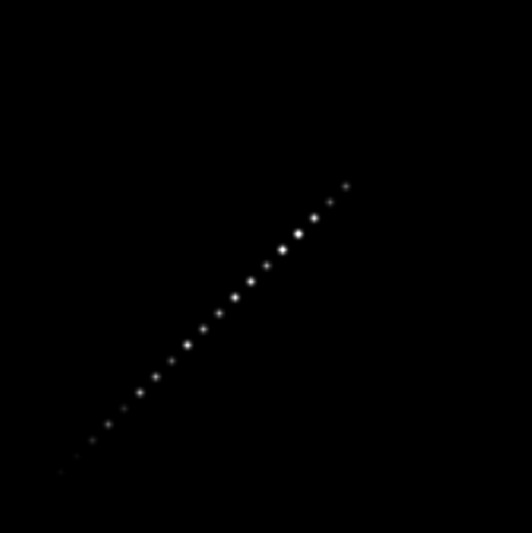} &
\includegraphics[width=0.2 \textwidth]{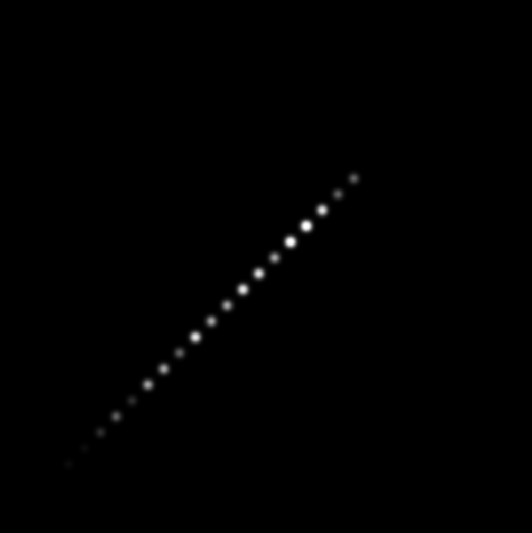} &
\includegraphics[width=0.2 \textwidth]{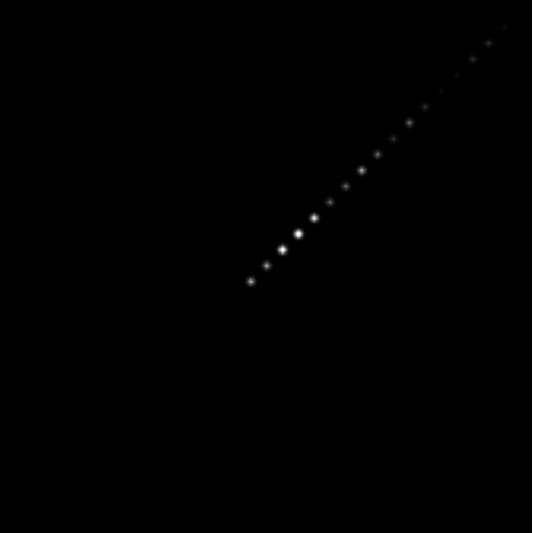} \\
{\tiny \#5 (1)*} &
{\tiny \#6 (1,1,2,3,2,4,4,3,5,4,4,5,5,4,6,6,5,3,3)*} &
{\tiny \#7 (1,1,2,3,2,4,4,3,5,4,4,5,5,4,6,6,5,3,3)*} &
{\tiny \#8 (4,4,6,6,5,3,3,4,3,2,3,2,1,1,2,2,1)*} \\
\hline
\includegraphics[width=0.2 \textwidth]{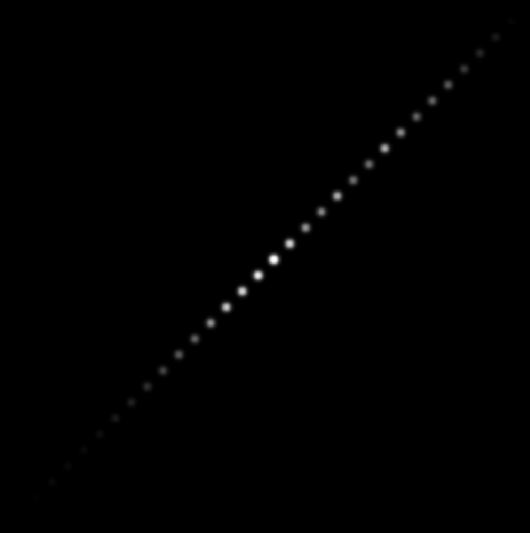} &
\includegraphics[width=0.2 \textwidth]{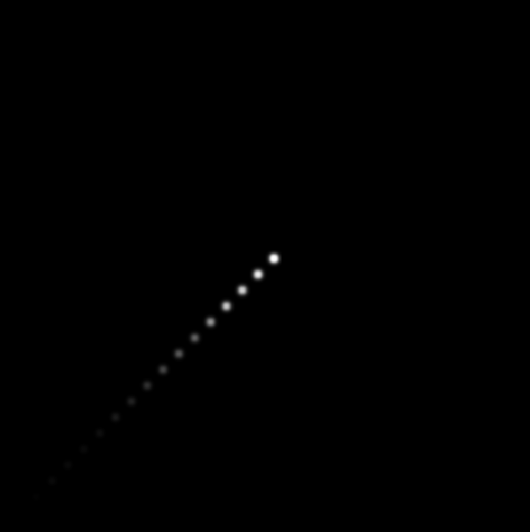} &
\includegraphics[width=0.2 \textwidth]{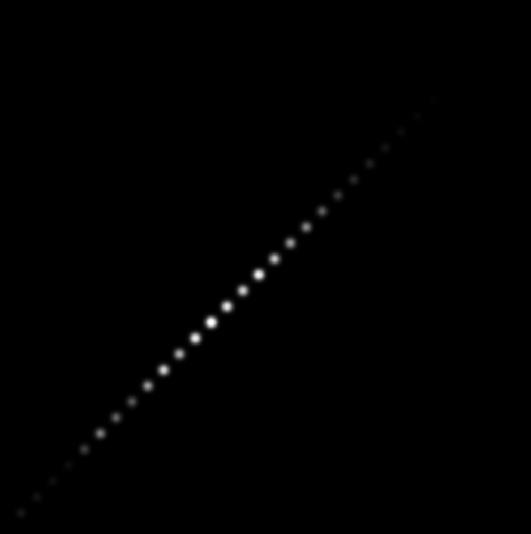} &
\includegraphics[width=0.2 \textwidth]{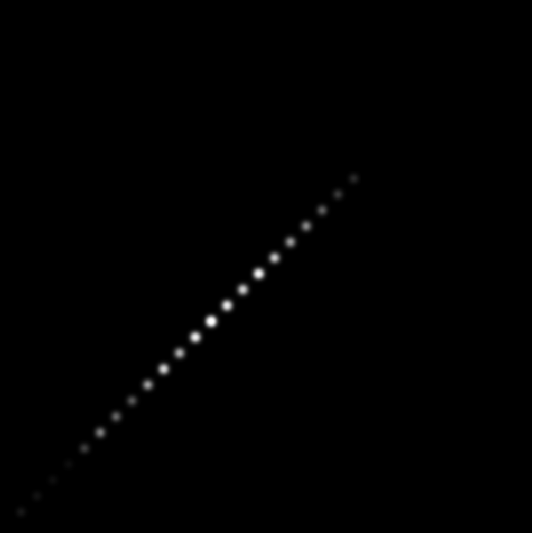} \\
{\tiny \#9 \begin{tabular}{c}(4,8,10,12,15,16,16,19,22,17,17,\\
22,19,20,25,32,28,25,24,20,16,\\
15,13,11,9,8,6,5,5,5,3,1)\end{tabular}} &
{\tiny \#10 (32,28,25,24,20,16,15,13,11,9,8,6,5,5,5,3,1)} &
{\tiny \#11 \begin{tabular}{c}(32,28,25,24,52,76,68,62,87,105,92,\\
115,133,114,102,121,100,86,79,66,\\
51,43,37,30,23,19,14,9,6,5,3,1)*\end{tabular}} &
{\tiny \#12 \begin{tabular}{c}(32,28,25,24,52,76,68,62,87,105,92,115, \\
133,114,102,121,100,86,79,66,51,43)*\end{tabular}} \\
\hline
\end{tabular} \\
{ } \\
\includegraphics[width=0.4 \textwidth]{mm16_12.PNG} \\
{\scriptsize \#13 (32,28,25,25,53,78,71,64,91,109,95,120,137,118,107,126,104,92,85,71,54,46)} \\
\end{tabular}
\caption{Images depict the digit encoding pattern after enumerated steps of the optical modular multiplication illustrated in Figure~\ref{lens_configuration}. Vector values are the integrated optical intensity for each digit. Those values are normalized and rounded to the nearest integer. An asterisk after a vector denotes values that will appear in reverse order in its corresponding image.}
\label{mod_mul_images}
\end{figure*}

The results of the computation are illustrated in Figure~\ref{mod_mul_images} that follows. Digit values printed below each image are the result of integrating the light intensity over the area of each digit and normalizing. The results are then rounded to the nearest integer. The maximum absolute error is $0.3419$ and the RMS error is $0.1442$, meaning that the analog result matches the correct integer solution. The main sources of error were found to be resolution and dynamic range. Specifically, the resolution in the Fourier transform plane needs to be sufficient for the pixel density to be at least twice the Nyquist frequency of the transform. This simulation used a 1 mm imaging area at each point-wise multiply and assumed a minimum pixel pitch of 1$\mu$m $\times$ 1$\mu$m. The dynamic range was assumed to be at least 12 stops (i.e., bits) or approximately 36 dB. The values $\mathcal{F}(M^*)$, $\mathcal{F}(m)$, $\mathcal{F}(1^*)$ and $\mathcal{F}(1)$ are precomputed inputs to the point-wise multiply unit. The performance bottleneck for this device will be the point-wise multiply by $\mathcal{F}(\bar{b})$ which must be performed in real time with each modular multiplication.

\section{Conclusions}
Optical techniques seem to be very promising in the future of high-performance computing.  The most significant area of future work will be to explore ways of performing a point-wise multiply optically.  If that can be done, this method may provide huge speed-ups over current technology.  Even if the point-wise multiply must be done with an auxiliary computer, the computation will be $O(n)$ instead of $O(n^2)$ thanks to the nearly instantaneous Fourier transform achieved optically.  If this path proves to be optimal, more work would need to be done to decrease the load for the electronic computer.  A lookup table approach is possible, as are methods of keeping field values at whole numbers.  Other future work might include finding the limits of how much information can be packed into a given area, how small the Fourier transform plane can be cropped, and how many operations can be done in series before the signal becomes unreadable.

Based on our simulation results it also appears feasible for an architecture similar to the one analyzed to accurately perform all-optical modular multiplication with greater precision than has been traditionally associated with analog computation~\cite{psaltis86}. Given our estimates of parameters such as size, resolution and dynamic range, such a device could be manufacturable from commercially available components. If these simulation results bear out in the lab then the ability to accurately perform 16-bit arithmetic could be a significant step forward for optical computation. A further contribution of this work is the demonstration of an approximate algorithm, implementable in optics, that is capable of correctly computing discrete modular multiplication. Future work should include further refinement to the simulation, optimization of the device configurations, particularly with regards to the lenses, and then  fabrication and characterization of an actual device.

\bibliographystyle{IEEEtran}

\end{document}